\documentclass[12pt]{article}
\usepackage{epsfig}
\usepackage{graphicx}
 \usepackage{latexsym,amsmath,tabularx,amssymb,bm}

\newcommand{\be}{\begin{equation}}
\newcommand{\ee}{\end{equation}}
\newcommand{\bea}{\begin{eqnarray}}
\newcommand{\eea}{\end{eqnarray}}

\def\simge{\mathrel{%
   \rlap{\raise 0.511ex \hbox{$>$}}{\lower 0.511ex \hbox{$\sim$}}}}
\def\simle{\mathrel{
   \rlap{\raise 0.511ex \hbox{$<$}}{\lower 0.511ex \hbox{$\sim$}}}}
\def\bigs{\mathrel{
   \rlap{\raise 0.531ex \hbox{$>$}}{\lower 0.531ex \hbox{$<$}}}}

\begin{document}
\begin{minipage}[t]{3cm}
 DESY-13-126\\
\end{minipage}
\vspace*{1.cm}
\begin{center}
\begin{Large}
Generalized Bootstrap Equations \\and possible implications for the NLO Odderon
\\[1cm]
\end{Large}
\vspace{0.5cm}
J. Bartels$^a$, G.P.Vacca$^b$  \\[1cm] 
$^a$ II. Institut f\"{u}r Theoretische Physik, Universit\"{a}t Hamburg,
Luruper Chaussee 149,\\D-22761 Hamburg, Germany\\
$^b$ INFN Sezione di Bologna, via Irnerio 46, I-40126 Bologna, Italy
\end{center}
\vskip15.0pt 
\noindent
{\bf Abstract:}\\
\noindent
We formulate and discuss generalized bootstrap equations in nonabelian gauge theories.
They are shown to hold in the leading logarithmic approximation. Since their validity  
is related to the self-consistency of the Steinmann relations for inelastic production amplitudes 
they can be  expected to be valid also in NLO. Specializing to the $N=4$ SYM, we show that    
the validity in NLO of these generalized bootstrap equations allows to find the NLO Odderon solution with 
intercept exactly at one, a result which is valid also for the planar limit of QCD.
 

\section{Introduction}
Bootstrap equations are among the most fundamental properties of the BFKL equation \cite{BFKL}. For the 
elastic $2 \to2$ scattering amplitude:
\be
T_{2\to2} = \frac{2s}{t} \int \frac{d\omega}{2\pi i} \,\,\xi\,\, (\frac{|s|}{\mu^2})^{\omega} F(\omega,t)
\ee
with the signature factor 
\be
\xi = - e^{-i\pi\omega} +\tau\,,
\ee 
(where the signature $\tau$ takes the value $\tau=+$ for the color singlet Pomeron channel and $\tau=-$ 
for the color octet exchange of the reggeized gluon) the well-known bootstrap property of the BFKL equation 
relates, via unitarity, for the color octet exchange channel  the energy discontinuity (i.e. the imaginary part) of  
$T_{2\to 2}$ to 
the (leading) real parts of $T_{2 \to n}$ production amplitudes\footnote{We define $disc_x f(x) =\frac{1}{2i} \left(f(x+i\epsilon) - f(x-i\epsilon)\right)$}: 
\be
disc_sT_{2 \to 2} = 2s \frac{\omega(t)}{t}  g^2 s^{\omega(t)} = \sum_n \int d\Omega_n |T_{2\to n}|^2,
\ee
where $\int d\Omega_n$ stands for the phase space integral of $n$ produced gluons, and 
\be
\alpha_g(t) = 1 + \omega(t)
\ee
denotes  the gluon trajectory.

This bootstrap equation was first derived in the leading logarithmic approximation \cite{BFKL}. In the attempt to 
generalize to NLO, Braun and Vacca \cite{Braun:1998zj,Braun:1999uz} introduced the notion of 'strong bootstrap':
one of the bootstrap properties is formulated as an eigenvalue condition of the two-reggeon Green's function  which shows the close connection between impact factors and the wave function of the reggeized gluon.
Conjecturing the validity of a set of strong bootstrap condition at NLO and making use of the NLO quark and gluon impact factor expressions they were able to compute the wave function of the reggeized gluon.  After the computation of the full NLO BFKL nonforward QCD kernel in the color octet channel, the fulfillment of these conditions were  explicitly proven by Fadin et al. \cite{Fadin:2002hz}. A first step towards generalizing the 
bootstrap condition to inelastic amplitudes (still on the basis of single energy discontinuities) was carried out 
in \cite{Bartels:2003jq}: apart from confirming the previous bootstrap relations, another bootstrap relation was formulated. 
This relation was finally proven in \cite{lastboot}. 

In this paper we turn to double energy discontinuities of $2 \to 3$ inelastic amplitudes and show that they lead to 
generalized bootstrap conditions. As a by-product we show that, if these generalized bootstrap conditions are valid 
also in NLO, they can be used to predict the NLO Odderon solution and its intercept.  

\section{Double discontinuities in LL}
\setcounter{equation}{0}
Consider the signatured $2\to3$ scattering amplitude in the double Regge limit. 
In accordance with the Steinmann relations, it can be written as a sum of the two terms:
\begin{center}
\epsfig{file=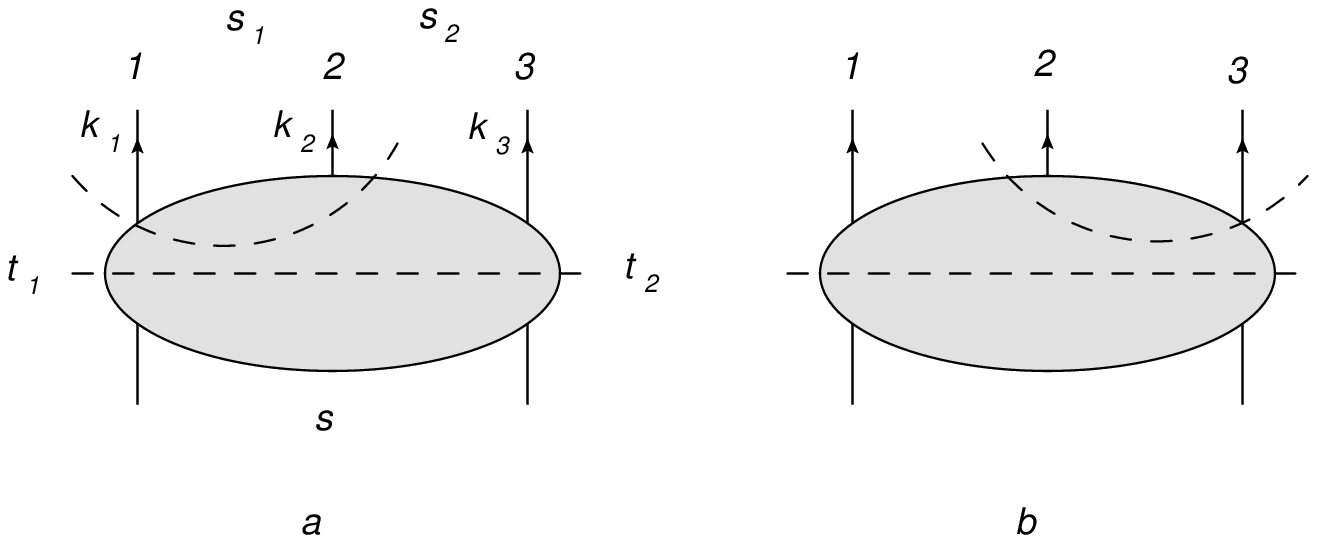,width=14cm,height=4cm}\vspace{1cm}\\
Fig.1: Decomposition for the $2\to3$ scattering amplitude\vspace{1cm}   
\end{center} 
\bea
T_{2\to3}=\frac{2s}{t_1t_2} \int \frac{d \omega_1}{2 \pi i}\int \frac{d \omega_2}{2 \pi i}
\Big[ \left(\frac{|s_1|}{\mu^2}\right)^{\omega_1-\omega_2} \left(\frac{|s|}{\mu^2}\right)^{\omega_2} \xi_{12}\xi_2 \frac{F_R(\omega_1,\omega_2;t_1,t_2;\eta)}{\Omega_{12}} \nonumber\\
+\left(\frac{|s_2|}{\mu^2}\right)^{\omega_2-\omega_1} \left(\frac{|s|}{\mu^2}\right)^{\omega_1} \xi_{21}\xi_1 \frac{F_L(\omega_1,\omega_2;t_1,t_2;\eta)}{\Omega_{21}}\Big]
\label{eq:decomp}
\eea
with
\be
s_1=(k_1+k_2)^2\,, \quad s_2=(k_2+k_3)^2\,, \quad \eta=\frac{s_1 s_2}{s}={\vec{k}}_{2}^2\,,
\ee
\be
\xi_i = e^{-i\pi \alpha(t_i) } + \tau_i  = -  e^{-i \pi \omega_i} + \tau_i\,,
\ee
\be
\xi_{ij} = e^{-i \pi(\omega_i -\omega_j)} + \tau_i \tau _j ,
\ee
and
\be
\Omega_i =\sin \pi \omega_i\,,\quad \Omega_{ij}=\sin \pi (\omega_i -\omega_j).
\ee

\subsection{The signatures $(\tau_1,\tau_2)=(-,-)$}
We first concentrate on the negative signature case: $\tau_1=\tau_2 =-$ with 
reggeized gluons in both $t$-channels. In the leading logarithmic approximation the amplitude is known to take the simple factorizing form:
\begin{center}
\epsfig{file=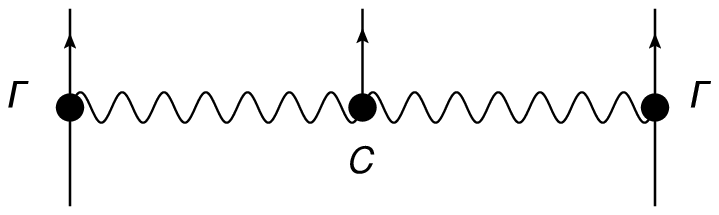,width=8cm,height=2cm}\vspace{1cm}\\
Fig.2: The leading log approximation of the $2\to3$ scattering amplitude\vspace{1cm}   
\end{center}  
\be
T_{2 \to 3} = \frac{2s}{t_1t_2}\Gamma(t_1) \left(\frac{|s_1|}{\mu^2}\right)^{\omega(t_1)} g C(q_1,q_2) \left(\frac{|s_2|}{\mu^2}\right)^{\omega(t_2)} \Gamma(t_2)\,,
\label{eq:LL}
\ee
where $\Gamma(t)$ and $C(q_1,q_2)$ denote the residue function and effective production 
vertex, resp. In the following we will show that the partial waves $F_R$ and $F_L$ can be determined 
from energy discontinuities and unitarity integrals. In particular, single and double 
discontinuities give the same answer, as a result of generalized bootstrap equations. 
Inserting the results for $F_R$ and $F_L$ into (\ref{eq:decomp}) we find agreement with   
(\ref{eq:LL}).

Before we start to discuss discontinuities it maybe useful to say a few words on how energy discontinuities fit into the hierarchy of leading log (LA), next-to-leading log (NLA), and next-next-leading log (NNLA) corrections to the $2\to3$ scattering amplitude. For the (-,-)-signature configuration, the amplitude in LA is real-valued, any single discontinuity is imaginary and belongs to NLA , whereas a double discontinuity is real-valued again and belongs to NNLA. These energy discontinuities result from the signature factors only. When considering the real-valued double discontinuity, this NNLA term should not be confused with higher order corrections of the partial waves $F_L$ or $F_R$: these have to be computed in a separate calculation. Our following discussion only addresses the energy discontinuities obtained from the signature factors, and we will study their compatibility with unitarity integrals.  

Both the single energy discontinuities in the subenergy $s_1$  and the double discontinuity in $s_1$ and $s$ determine $F_R$. From (\ref{eq:decomp}) we find that 
\be 
disc_{s_1} T_{2 \to 3} =- \frac{2s}{t_1t_2} \int \frac{d \omega_1}{2 \pi i}\int \frac{d \omega_2}{2 \pi i}
 \left(\frac{|s_1|}{\mu^2}\right)^{\omega_1-\omega_2} \left(\frac{|s|}{\mu^2}\right)^{\omega_2} \xi_2 F_R(\omega_1,\omega_2;t_1,t_2;\eta)
 \label{eq:s1discontinuity}
\ee
and 
\be
disc_{s_1}disc_{s}T_{2 \to 3}= \frac{2s}{t_1t_2} \int \frac{d \omega_1}{2 \pi i}\int \frac{d \omega_2}{2 \pi i}
\left(\frac{|s_1|}{\mu^2}\right)^{\omega_1-\omega_2} \left(\frac{|s|}{\mu^2}\right)^{\omega_2}  
\Omega_2
F_R(\omega_1,\omega_2;t_1,t_2;\eta) 
\label{eq:ss2discontinuity}
\ee
both determine $F_R$. The diagrams corresponding to the single and double 
discontinuities are shown in Fig.2a and 2b, resp.
\begin{center}
\epsfig{file=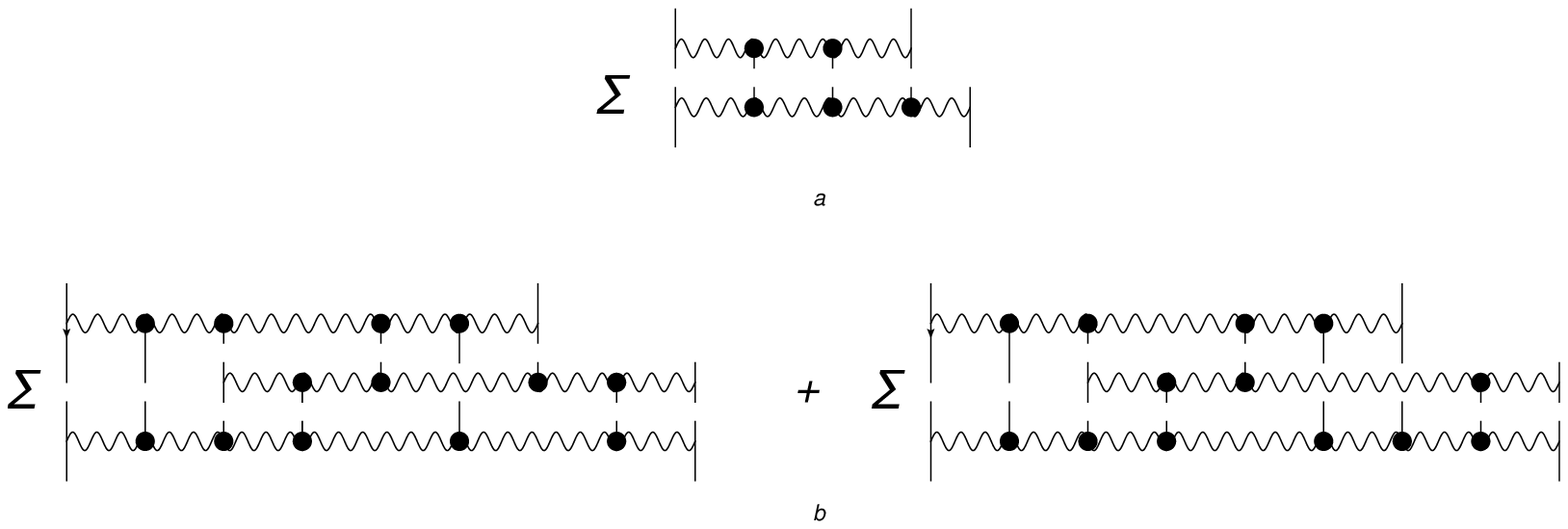,width=12cm,height=5cm}\vspace{1cm}\\
Fig.3: Single and double energy discontinuities of the $2\to3$ scattering amplitude.\\
Dots denote the effective BFKL production vertex.
\vspace{1cm}   
\end{center}  

When computing the unitary integrals illustrated in Fig.3a, it is convenient to transform  
into the CM system of the outgoing particles 1 and 2, using helicity conservation of the elastic scattering of particle 2, performing the 'multiplication' and then moving back into the overall CM system. Alternatively, one can use left and right gauges for the polarization vectors of the produced gluon. The result for the production vertices $_2V_1$ and $_2V_2$ are illustrated 
in Fig.4a and 4b, resp..
\begin{center}
\vspace{1cm}
\epsfig{file=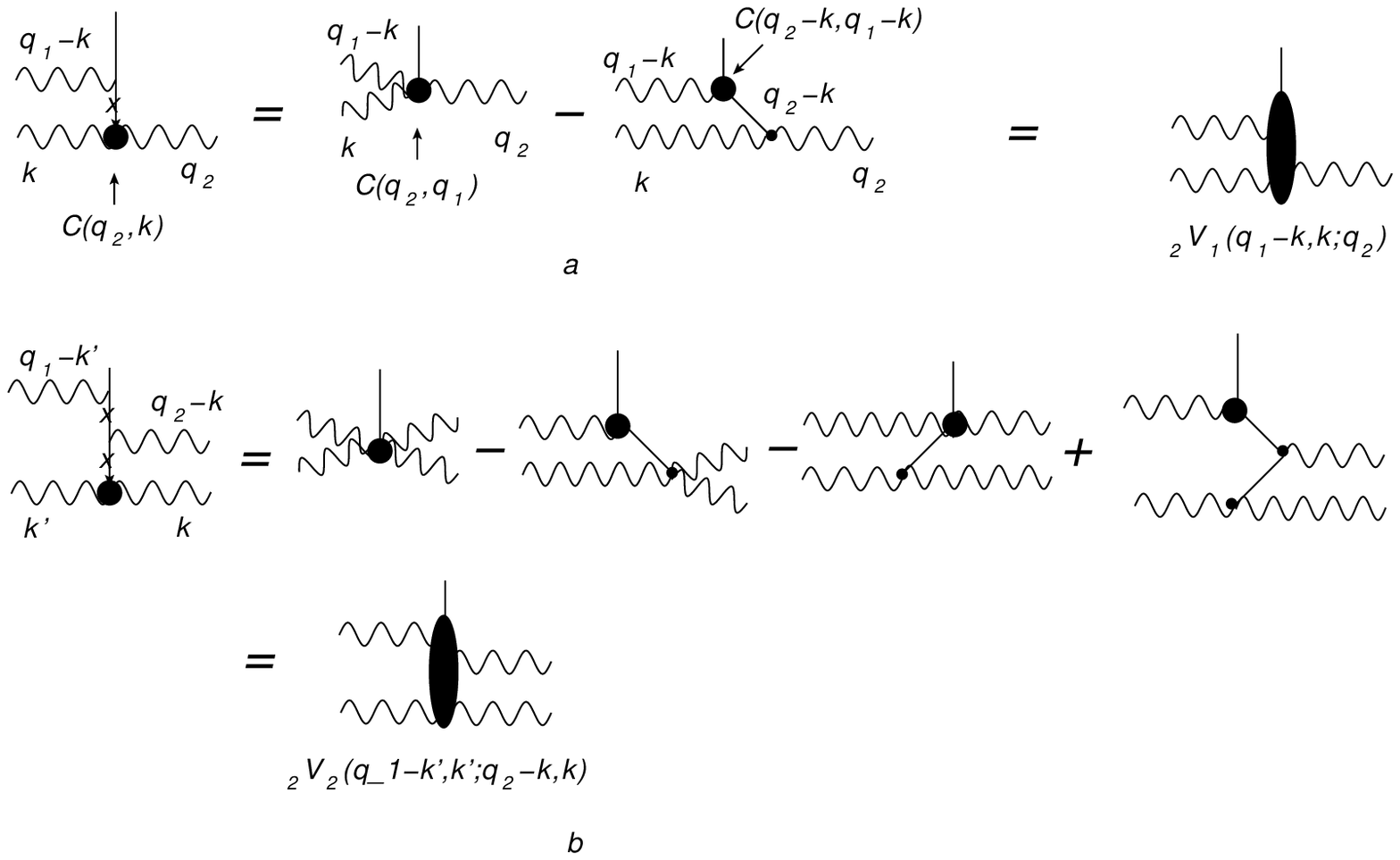,width=16cm,height=8cm}\vspace{1cm}\\
Fig.4. Production vertices: a) the RRPR production vertex $_2V_1$ , b) the RRPRR production vertex $_2V_2$\\
black dots denote effective production vertices, straight line the gluon propagator $1/k^2$.  
\vspace{1cm}   
\end{center}        

Next we define (amputated) BFKL amplitudes for two reggeized gluons, 
$D_2^{a_1a_2}(k_1,k_2;q|\omega)$, where $k_1+k_2=q$, $a_1,a_2$ are the color indices of the reggeized gluons, and  we suppress the color 
labels of the external particles\footnote{The notion 'amputated' refers to the transverse momentum propagators of the reggeized gluons. Our function $D_2(k_1,k_2,q;\omega)$  includes, for the outgoing reggeons, a reggeon 
denominator  $1/(\omega - \omega(k_1)-\omega(k_2))$.}. 
For the color octet channel we factor out the overall color tensor and define the 
two-reggeon amplitude $D_2^{(8_A)}$ which satisfies the integral equation (rhs of Fig.5):
\be
\omega D_2^{(8_A)}(k_1,k_2;q|\omega)=D_{(2; 0}^{(8_A)} (k_1,k_2;q) + \left( \big[ K_{8r} +\omega(k_1) + \omega(k_2) \big] \otimes 
D_2^{(8_A)} \right) (k_1,k_2;q|\omega)\,,
\ee 
where $K_{8r}$ is the real emission part of the color octet BFKL kernel (i.e. it contains the color factor $\frac{N}{2}$ and,
in LO, it equals $\frac{1}{2} K_{1r})$.   The integration in transverse momentum variables is denoted with $\otimes=\int d^2k'_1 d^2k'_2 \,\delta^{(2)}(q\!-k'_1\!-k'_2)1/(k'_1{}^2 k'_2{}^2)$.  
This equation has the familiar bootstrap solution  
\be
D_2^{(8_A)}(k_1,k_2;q|\omega)=\Gamma(t) \frac{1}{\omega - \omega(q)} g
\label{eq:2bootstrap}
\ee
with $-q^2=t$. In obtaining (\ref{eq:2bootstrap}) it was essential that the inhomgeneous term, 
$D_{2; 0}^{(8_A)} (k_1,k_2;q)$ is point-like, i.e it depends only on $k_1+k_2=q$ and not separately on $k_1$ and $k_2$. 
 \begin{center}
 \vspace{1cm}
\epsfig{file=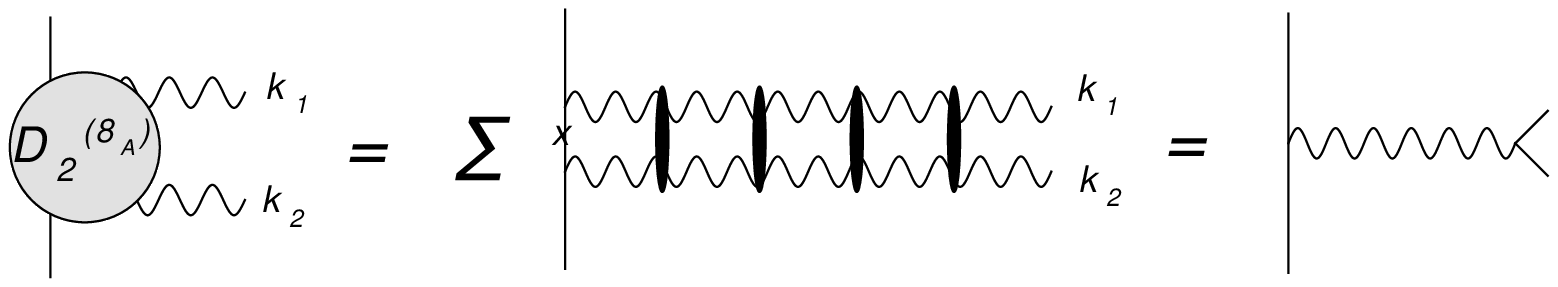,width=10cm,height=2cm}\vspace{1cm}\\
Fig.5: bootstrap equation of the color octet Green's function  
\vspace{1cm}   
\end{center}  
In order to obtain the left hand part of Fig.3a, we convolute the two-reggeon amplitude $D_{2}^{(8_A)} (k_1,k_2;q_1|\omega_1)$ (where $k_1+k_2=q_1$) with the production vertex $_2V_1(k_1,k_2;q_2)$ and use the bootstrap solution 
(\ref{eq:2bootstrap}). After some algebra we find the following expression for the partial wave $F_1$: 
\be
F_1= \Gamma(t_1) \frac{1}{\omega_1 - \omega(q_1)}
 g\pi C(q_2,q_1) \left(\frac{1}{2}
(\omega(q_1)-\omega(q_2)) -\frac{a}{2} (\ln(\frac{\eta}{\mu^2} -\frac{1}{\epsilon})\right)
 \frac{1}{\omega_2 - \omega(q_2)}  \Gamma(t_2).
\ee

Next we show that the same result also follows from the double discontinuity. Starting from the left hand part of 
Fig.3b, we first define
the (amputated) three-reggeon amplitudes $D_{3}^{a_1a_2a_3}(k_1,k_2,k_3;q_1|\omega)$ where $k_1+k_2+k_3=q_1$, and $a_1,a_2,a_3$ are  
the color indices of the three reggeons. Together with the two-reggeon amplitudes, they satisfy the following 
coupled integral equations:
\be
\omega D_2^{a_1a_2}(k_1,k_2;q_1|\omega) =\nonumber
\ee
\be 
D_{2;0}^{a_1a_2}(k_1,k_2;q_1) + \left( \big[ K_r^{\{a_i\}} + \omega(k_1) +\omega(k_2) \big]\otimes  D_{2}^{\{a_i\}} 
\right)^{a_1a_2}(k_1,k_2;q_1|\omega)
\label{eq:D2equation}
\ee
\be
\omega D_3^{a_1a_2a_3}(k_1,k_2,k_3;q_1|\omega) =\nonumber
\ee
\bea D_{3;0}^{a_1a_2a_3}(k_1,k_2,k_3;q_1) + 
\left( K_{2\to3}^{\{a_i\}} \otimes  D_{2}^{\{a_i\}} 
\right)^{a_1a_2a_3}(k_1,k_2,k_3;q_1|\omega)+\nonumber\\
\left( \big[ \sum_{ij} K_{r;ij}^{\{a_i\}} + \omega(k_1) +\omega(k_2) + \omega(k_3) \big] \otimes 
D_3^{a_i} \right)^{a_1a_2a_3} (k_1,k_2,k_3;q_1|\omega),
\hspace{1cm}
\label{eq:D3equation}
\eea
where the color superscripts denote the color structure, and lower subscripts refer to reggeon lines. In the second equation, $k_1+k_2+k_3=q_1$.
$K_{r;23}^{\{a_i\}}$ is the real emission part of the 
BFKL kernel acting between the reggeons '2' and '3', and it includes the structure constants $f^{abc}$:
\bea
K_{r}^{\{a_i\}}(q_1,q_2;q'_1,q'_2) &=& f^{a_1a'_1l}f^{la_2a'_2}K_r(q_1,q_2;q'_1,q'_2)\nonumber\\
\frac{1}{g^2}K_r &=& \left(-q^2+ \frac{q_1^2{q'}_2^2+{q'}_1^2q_2^2}{(q_1-q'_1)^2}
\right)\,.
\eea
The transition kernel  $K_{2\to3}^{\{a_i\}} = K_{2\to3} f^{a_1a'_1l}\,f^{l a'_2 m}\,f^{a'_3a_3m}$ has the form:
\bea
\frac{1}{g^3} K_{2\to 3}(q'_1,q'_2,q'_3;q_1,q_3)=- q^2 -\frac{q'^{2}_2 q_1^2 q_3^2}{(q'_1\!-\!q_1)^2(q'_3\!-\!q_3)^2}
+\frac{(q'_1\!+\!q'_2)^2 q_3^2}{(q'_3\!-\!q_3)^2}+\frac{(q'_2\!+\!q'_3)^2q_1^2 }{(q'_1\!-\!q_1)^2}
\label{eq:K2to3}
\eea
with $q=q'_1\!+\!q'_2\!+\!q'_3$. We would like to stress that this transition kernel is symmetric under $q_1 \leftrightarrow q_3,
\, q'_1 \leftrightarrow q'_3$;
in Fig. 3b, it is obtained as the product of two effective production vertices above and below and an elastic rescattering vertex in between.

We consider the special case where we antisymmetrize in the reggeons with momenta $k_2$ and $k_3$:
\bea
A_{23}*D_3^{a_1a_2a_3}(k_1,k_2,k_3;q_1|\omega) = &\nonumber\\
\frac{1}{2} \left( D_3^{a_1a_2a_3}(k_1,k_2,k_3;q_1|\omega)
- D_3^{a_1a_3a_2}(k_1,k_3,k_2;q_1|\omega)\right)   
\eea
(this defines the operator $A_{23}$ which antisymmetrizes in the two lower legs with momenta $k_2$ and $k_3$ and color $a_2$ and $a_3$).
This function satisfies the equation
\be
\omega A_{23}*D_3^{a_1a_2a_3}(k_1,k_2,k_3;q_1|\omega) =\nonumber
\ee
\bea A_{23}* D_{3;0}^{a_1a_2a_3}(k_1,k_2,k_3;q_1) + 
\left( (A_{23}* K_{2\to3}^{\{a_i\}}) \otimes  D_{2}^{\{a_i\}} 
\right)^{a_1a_2a_3}(k_1,k_2,k_3;q_1|\omega)+\nonumber\\
\left( \big[ \sum_{ij} K_{r;ij}^{\{a_i\}} + \omega(k_1) +\omega(k_2) + \omega(k_3) \big] \otimes 
A_{23}*D_3^{a_i} \right)^{a_1a_2a_3} (k_1,k_2,k_3;q_1|\omega)\,.
\hspace{1cm}
\label{eq:antisymD3equation}
\eea 

Provided the corresponding inhomogeneous term $A_{23}*D_{3;0}^{a_1a_2a_3}(k_1,k_2,k_3;q_1)$ is pointlike in the 
pair of momenta $k_2$ and $k_3$ (i.e. it depends only upon the sum $k_2+k_3=k_{23}$) and equals 
\be 
A_{23}*D_{3;0}^{a_1a_2a_3}(k_1,k_2,k_3;q_1) = D_{2;0}^{a_1 c }(k_1,k_{23};q_1) \frac{g}{2} f^{c a_2a_3},
\label{eq:bootstrapimpactfactor} 
\ee
the equation for  $A_{23}*D_3^{a_1a_2a_3}(k_1,k_2,k_3;q_1;\omega)$ can be solved. One obtains:
\be
A_{23}*D_3^{a_1a_2a_3}(k_1,k_2,k_3;q_1|\omega) = D_2^{a_1c }(k_1,k_{23};q_1|\omega) \frac{g}{2} f^{c a_2a_3}\,.
\label{eq:3bootstrap}
\ee
\begin{center}
\vspace{1cm}
\epsfig{file=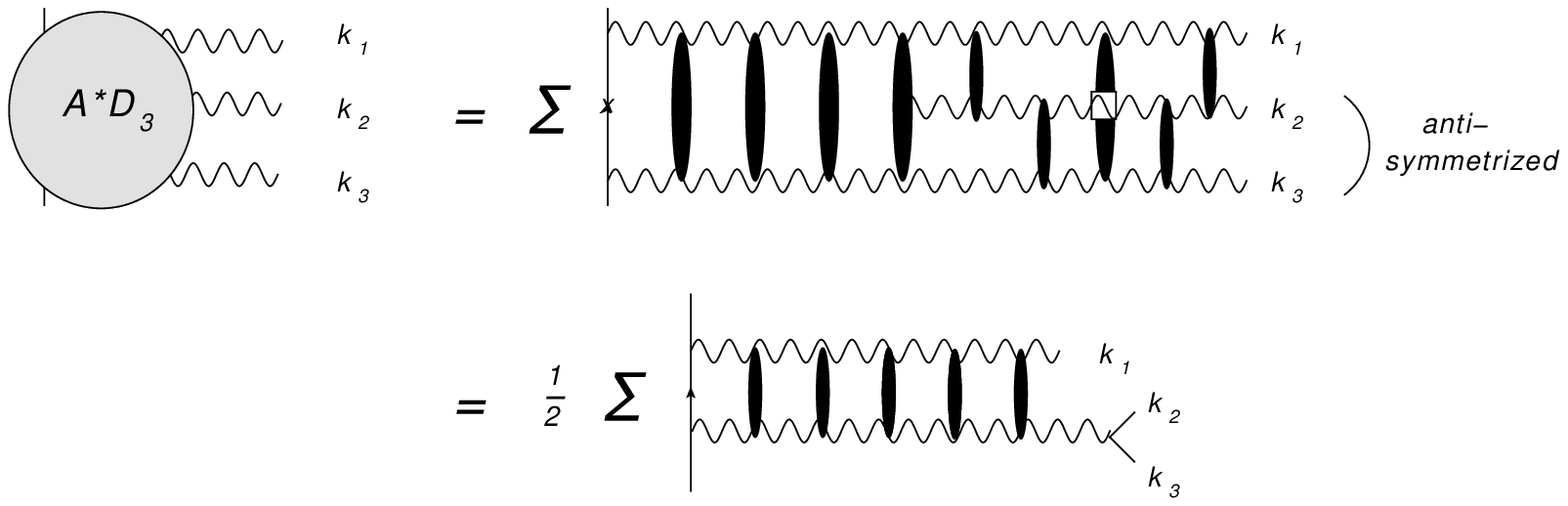,width=15cm,height=5cm}\vspace{1cm}\\
Fig.6: the bootstrap relation (\ref{eq:3bootstrap}) 
\vspace{1cm} 
\end{center}
In order to verify this solution, one inserts the solution into the integral equation for $A_{23}*D_3^{a_1a_2a_3}(k_1,k_2,k_2;q_1|\omega)$, (\ref{eq:antisymD3equation}),  and makes use of the following relation between the $2\to3$ kernel and the BFKL kernel \footnote{In all our figures, the kernels are acting on wave functions to the left (sign of arrows) whereas 
in our formulae operators are acting on wave functions on the right} :
\be
[K_{r;1'2'}, S_{2}] =- K_{2\to3},
\label{eq:commutator}
\ee
where the operator $S_{2}$ denotes the splitting of line 2. We illustrate this equation in Fig.7a:
 \begin{center}
 \vspace{1cm}
\epsfig{file=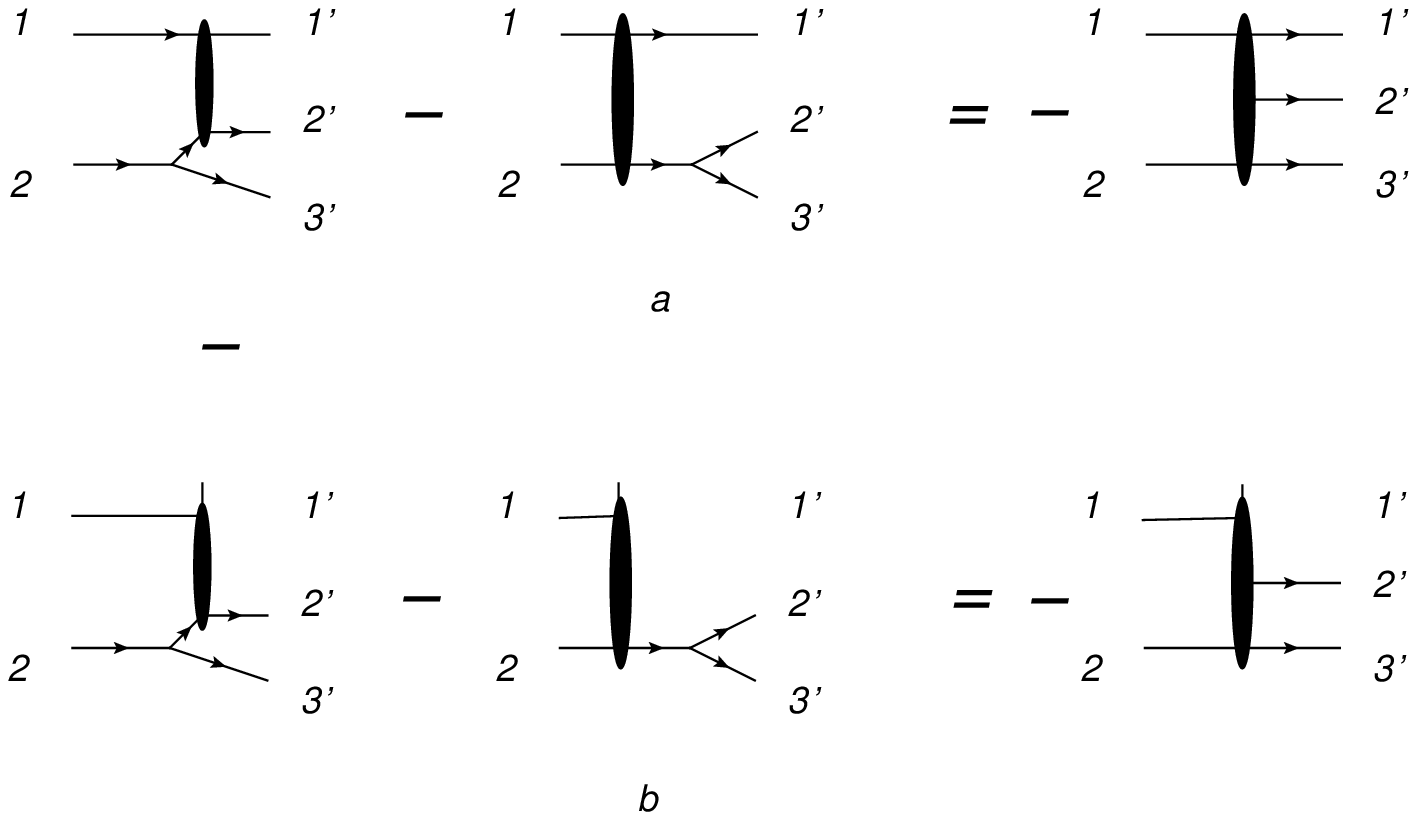,width=12cm,height=5cm}\vspace{1cm}\\
Fig.7: commutator relations:\\ a) eq.(\ref{eq:commutator}) for the real emission part of the BFKL kernel \\b) for the RRPR production vertex. \\Both identities hold for the momentum structures $K_r$ and $K_{2\to3}$, i.e. color has been stripped off.
\vspace{1cm}   
\end{center} 
A similar equation holds for the RRPR production vertex, $_2V_1$. We show this in Fig.7b.

The bootstrap condition (\ref{eq:3bootstrap}) represents a generalization of  the well-known bootstrap property of the two-reggeon Green's function (\ref{eq:2bootstrap}). Relations of this kind where first derived in 
~\cite{Bartels:1994jj, Braun:1997nu, Bartels:1999aw}. Note that, for this solution, we  have not projected on any total color quantum  number in the $t_1$ channel. In particular, this relation can be applied both to the color singlet and to the color octet channel.   

Let us now apply this result to the double discontinuity in Fig.3b.
In order to obtain, from the three-reggeon amplitude $A_{23}*D_3^{a_1a_2a_3}(k_1,k_2,k_3;\omega_1)$, the left hand part of Fig.3b, we have to convolute with the RRPR vertex and add the convolution of the two-reggeon Green's 
function with the RRPRR vertex (Fig.8):
\bea
_2V_1(q_1-k'-k,k'; q_2-k) \otimes A*D_3^{\{a_i\}}(q_1-k'-k,k',k;\omega_1)\nonumber \\
+_2V_1(q_1-q_2-k'+k,k'; k) \otimes A*D_3^{\{a_i\}}(q_1-q_2-k'+k,q_2-k,k';\omega_1)\nonumber\\
+  A* _2V_2(q_1-k',k';q_2-k,k) \otimes D_2^{\{a_i\}}(q_1-k',k';\omega_1) \nonumber\\
=\left(_2V_1(q_1-k',k';q_2) \frac{1}{2}gf^{ca_2a_3}\right) \otimes D_2^{\{a_i\}}(q_1-k',k';\omega_1)\,.
\label{eq:doubledisc}
\eea
 \begin{center}
 \vspace{1cm}
\epsfig{file=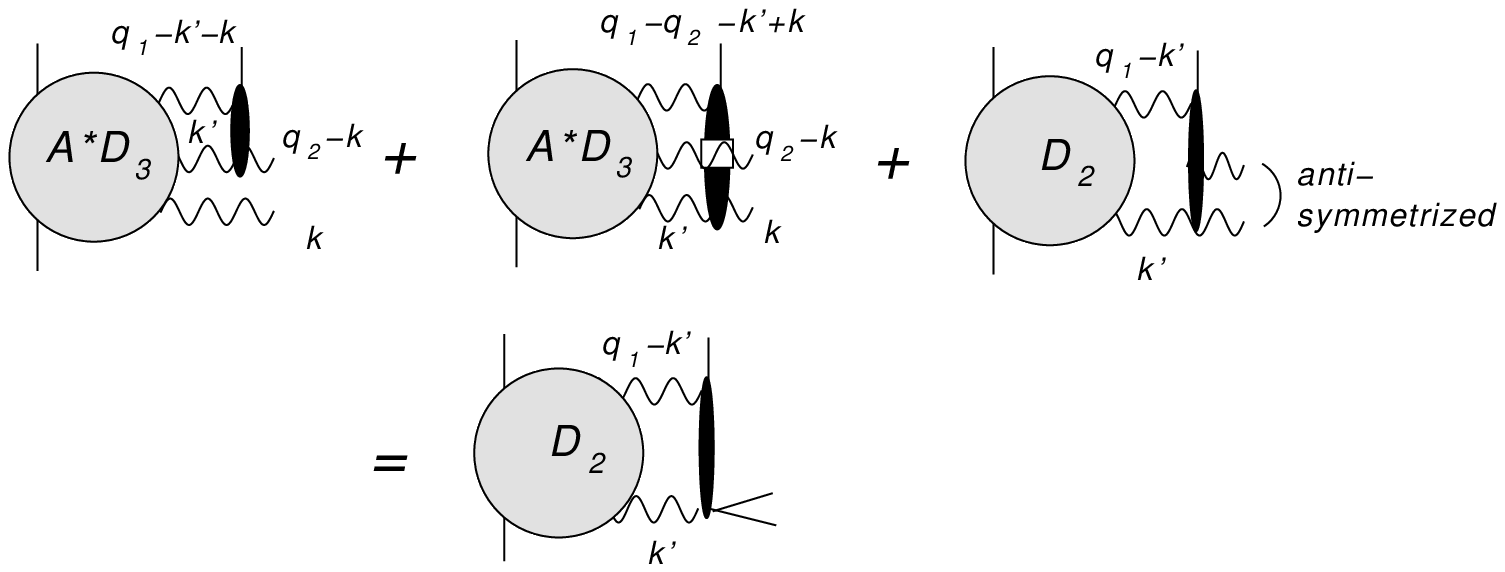,width=16cm,height=6cm}\vspace{1cm}\\
Fig.8: illustration of (\ref{eq:doubledisc})
\vspace{1cm}   
\end{center} 
Applying the bootstrap equation (\ref{eq:3bootstrap}) to $A_{23}*D_3$, and using the commutator shown in Fig.7b, we obtain the solution illustrated in Fig.8. Finally, we have to attach, on the rhs of Fig.8, the two-reggeon Green's function with 
the solution \ref{eq:2bootstrap}). It leads to the single discontinuity in Fig.3a, multiplied with $\omega(q_2)$, in full 
agreement with (\ref{eq:ss2discontinuity}). We have thus shown that, thanks to the bootstrap condition  (\ref{eq:3bootstrap}), single and double energy discontinuities are consistent and lead to the same answer for the partial 
wave $F_1$. 
 
Finally, an analogous discussion also applies to the second term in our starting decomposition (\ref{eq:decomp}),
to the partial wave $F_2$. One finds: 
 \be
F_2= \Gamma(t_1) \frac{1}{\omega_1 - \omega(q_1)}
 g\pi C(q_2,q_1) \left(\frac{1}{2}
(\omega(q_2)-\omega(q_1)) -\frac{a}{2} (\ln(\frac{\eta}{\mu^2} -\frac{1}{\epsilon})\right)
 \frac{1}{\omega_2 - \omega(q_2)}  \Gamma(t_2)\,.
\ee 
Inserting $F_1$ and $F_2$ into  (\ref{eq:decomp}) one easily verifies (\ref{eq:LL}).

We  conclude this section by a comment on the generalized bootstrap relation (\ref{eq:3bootstrap}).
As we have mentioned, the validity of this relation is related to the commutator (\ref {eq:commutator}) which
expresses the connection between the BFKL kernel and the $2\to 3 $ kernel. This commutator relation is easily verified 
from  by direct calculation from the analytic form of $K_r$ and $K_{2 \to 3}$. 
Alternatively, one can invert the argument and 
ask whether this commutator relation can be derived from the validity of the bootstrap condition (\ref{eq:3bootstrap}).
Starting from the integral equation for  $A_{23}*D_3^{a_1a_2a_3}(k_1,k_2,k_3;q_1;\omega)$, (\ref{eq:antisymD3equation}), and inserting the bootstrap solution (\ref{eq:3bootstrap}), one concludes that, 
on order that  (\ref{eq:3bootstrap}) solves the equation, the following (slightly weaker) condition must hold:
\begin{center}
\vspace{1cm}
\epsfig{file=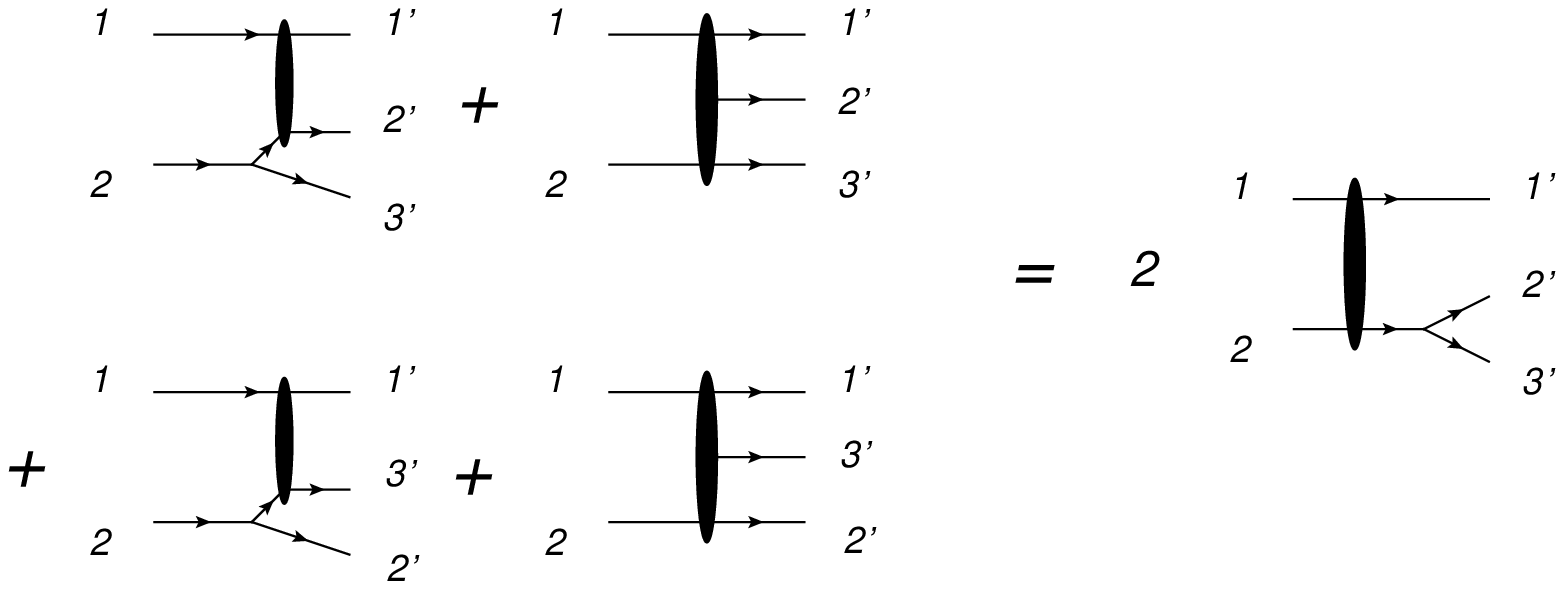,width=16cm,height=6cm}\vspace{1cm}\\
Fig.9: The modified commutator equation, derived from the validity of the bootstrap condition  (\ref{eq:3bootstrap}).
The kernels are the momentum space expressions $K_r$ and $K_{2 \to 3}$, i.e. color has has been stripped off.
\vspace{1cm}   
\end{center}
We write this equation in the following form: 
\be
K_{r;12} \otimes_{12}  S_{q_2} \otimes_{2} + K_{2 \to 3}(1'2'3') +K_{r;13} \otimes_{13} S_{q_2} \otimes_{2} + K_{2\to3} (1'3'2')  =2S_{q'_2} \otimes_{2}  K_{r;12} \,.
\label{eq:LOboot}
\ee
Later on we will show that this relation enables us to derive the LO Odderon solution: {\it the 
Odderon wave function and intercept can be derived from of the validity of the generalized bootstrap condition}. 
Generalizing the validity of (\ref{eq:3bootstrap}) to NLO, the same line of arguments allows to address also the NLO Odderon solution.   

\subsection{The signatures $(\tau_1,\tau_2)=(+,-)$}

Let us apply the same argument to another signature configuration, $(+,-)$: now we have the BFKL Pomeron in the 
$t_1$-channel and the reggeized gluon in the $t_2$ channel. The full $2 \to3$ scattering amplitude has again 
the representation (\ref{eq:decomp}), but now, in the leading logarithmic representation, only the first term 
contributes (the second one is of higher order), and it is directly proportional to the discontinuity in $s_1$ (Fig.3a).
The consistency between the single discontinuity in $s_1$ and the double discontinuity in $s_1$ and $s$ requires,
again, a relation between Figs.3a and b. As we have shown above, the bootstrap condition (\ref{eq:3bootstrap}) 
is valid both for the octet and the singlet representations in the $t_1$ channels. Therefore, our derivation of the 
relation between single and double energy discontinuities is the same as before: Fig.8 holds for singlet and octet 
representations in the $t_1$ channel.     

\section{NLO considerations}  
\setcounter{equation}{0}

Let us now turn to NLO. Our previous discussion has shown that the validity of the bootstrap condition (\ref{eq:3bootstrap}) is closely connected with the ansatz (\ref{eq:decomp}) for the $2 \to 3$ scattering amplitude and its 
double energy discontinuities. Similar to the familiar bootstrap equation, which was connected with the single 
energy discontinuity of the $2 \to 2$ scattering amplitude and which was shown to be valid in LO and NLO,    
it is plausible to expect that also (\ref{eq:3bootstrap}) remains valid in NLO. In the following we 
assume that this is the case, and we investigate the consequences.
We now concentrate on the color singlet $t_1$-channel (i.e. the signature configuration  $(\tau_1,\tau_2)=(+,-)$)
which has an overall color coefficient $f^{a_1a_2a_3}$ for the three-reggeon amplitude $D_3$.

Let us first see how the integral equations (\ref{eq:D2equation}) and (\ref{eq:D3equation}) are modified in NLO\footnote{From now on we will use superscripts $(0)$ and $(1)$ for distinguishing between LO and NLO quantities.}:\\
\noindent
1) the gluon trajectory function is to be taken in  NLO\\
2) in the equation for $D_2$,  we need the color singlet NLO impact factor $D_{2;0}^{(1)}$.  It is known for several cases 
(e.g. the quark and gluon impact factors and  the virtual photon impact factor).  
The NLO BFKL kernel in the color singlet representation is also known.\\
3) In the equation for $D_3$, we need the NLO color singlet  impact factor $D_{3;0}(k_1,k_2,k_3)^{(1})$ proportional to $f_{abc}$. This impact factor, when antisymmetrized in the two gluons with momenta $k_2$ and $k_3$, has to fulfill the strong bootstrap condition:  
\be
A_{23}*D_{3;0}^{(1)}(k_1,k_2,k_3;q_1) = \frac{1}{2} \left(D_{2;0}^{(1)}(k_1,k_2+k_3;q_1) \psi_{8}^{(0)}(k_2,k_3) 
                                           + D_{2;0}^{(0)}(k_1,k_2+k_3;q_1) \psi_{8}^{(1)}(k_2,k_3)\right)\,,
\ee
where $\psi_{8}^{(0)}=g$ and $\psi_{8}^{(1)}(k_2,k_3)$ denote the wave function of the reggeized gluon in LO and NLO, resp.. \\
4) In (\ref{eq:D3equation}), the NLO BFKL kernel acting in the three reggeon state is in the antisymmetric color octet representation (f-type).\\
5) the $2\to3$ kernel (cf.(\ref{eq:K2to3})) is to be computed in NLO (not known yet): by construction, it must be 
symmetric under the exchange: 
\begin{center}
(reggeons $1$ and $1'$) $\leftrightarrow$ (reggeons $3$ and $3'$) . 
\end{center}
6) In (\ref{eq:D3equation}) the integral kernel of the three-reggeon state receives a new contribution, a $3 \to 3$ 
kernel $K_{3\to3}(q_1,q_2,q_3; q'_1,q'_2,q'_3)$. It arises from inserting, into the double energy discontinuity, a production vertex which is computed in 
quasiregge kinematics (QMRK). We illustrate the three contributions in the following Fig.10:
\begin{center}
\vspace{1cm}
\epsfig{file=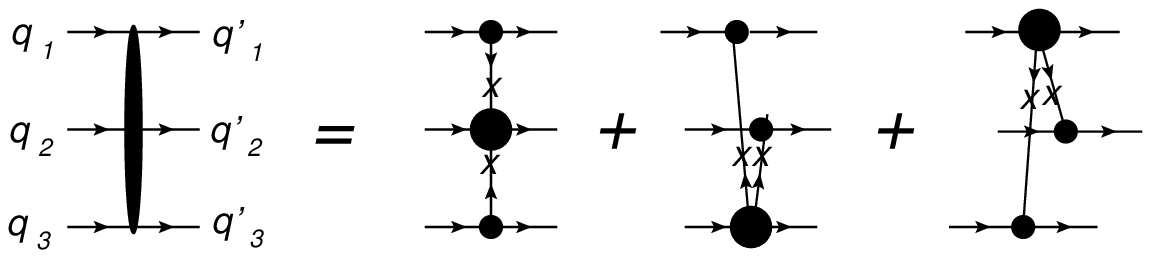,width=12cm,height=3cm}\vspace{1cm}\\
Fig.10: The momentum structure of the $3 \to 3$ kernel in the NLO integral equation for $D_3$ and $A_{23}*D_3$. 
The small blobs denote LO BFKL effective production vertices, the big blobs RPPR vertices in QMRK.
Vertical gluon lines are on-shell. 
\vspace{1cm}   
\end{center}
The computation, most conveniently, uses Lipatov's effective action and follows the arguments given in \cite{Bartels:2012sw}.
The detailed form of this kernel will not be presented here. We only mention the following symmetry properties:
\be
K_{3\to3}(q_1,q_2,q_3; q'_1,q'_2,q'_3)= K_{3\to3}(q_1,q_3,q_2; q'_1,q'_3,q'_2)
\ee
and 
\be
K_{3\to3}(q_1,q_2,q_3; q'_1,q'_2,q'_3)= K_{3\to3}(q_3,q_2,q_1; q'_3,q'_2,q'_1)\,.
\ee
With these modifications we can write down the NLO analogue of the integral equation for $A_{23}*D_3$, (\ref{eq:antisymD3equation}): apart from the NLO corrections to the kernels we have the new $3 \to 3$ kernel. 

Let us now follow the argument given at the end of section 2.1: we assume that this NLO integral equation for $A_{23}*D_{3}^{(1)}(k_1,k_2,k_3;q_1|\omega)$ has the solution
\be
A_{23}*D_{3}^{(1)}(k_1,k_2,k_3;q_1|\omega) = \frac{1}{2}\left( D_{2}^{(1)}(k_1,k_{23};q_1|\omega) \psi_8^{(0)}(k_2,k_3) +  D_{2}^{(0)}(k_1,k_{23};q_1|\omega) \psi_8^{(1)}(k_2,k_3)\right),
\ee
where, for simplicity, we have supressed the color factors.  
We insert this solution into the integral equations, and derive the condition under which the integral equation is 
satisfied. In analogy with Fig.9 we find:
\begin{center}
\vspace{1cm}
\epsfig{file=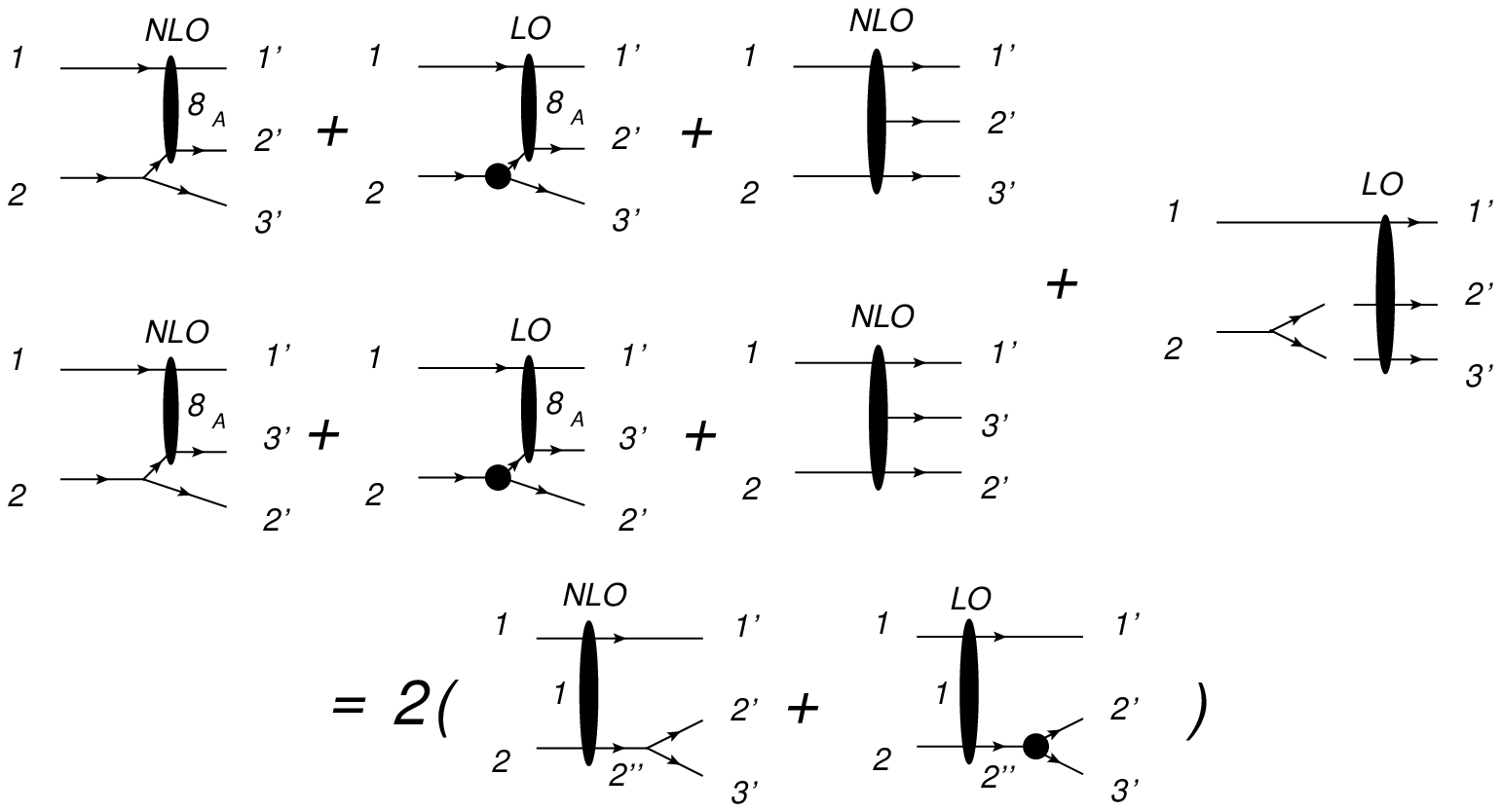,width=16cm,height=6cm}\vspace{1cm}\\
Fig.11: NLO analogue of Fig.9 (see text).\\ 
\vspace{1cm}   
\end{center}  
In Fig.11 we have stripped off all color factors in the following way. From the left we have projected onto the  color singlet state by contracting the color labels of the two gluons. The BFKL kernels (containing  only the real emission parts) are normalized to the projectors $P_{8_A}$ or $P_{1}$. The $3\to3$ kernel has been contracted, from the left, with the 
$f_{abc}$ tensor. In this way, all terms are proportional to the color tensor $\frac{1}{2} f_{a_1a_2a_3}$.  
The black blob denotes the NLO wave function of the 
reggeized gluon, $\psi_8^{(1)}$.  We give here a formula which encodes the relation of Fig. 11. After moving all terms
to the lhs we obtain:
\bea
&K_{8r;1'2'}^{(1)}\psi_8^{(0)}S_{2}+  K_{8r;1'2'}^{(0)} \psi_8^{(1)}S_{2} +K_{2\to3}^{(1)}(1',2',3') \nonumber\\
&+K_{8r;1'3'}^{(1)} \psi_8^{(0)} S_{2} +  K_{8r;1'3'}^{(0)} \psi_8^{(1)}S_{2} +K_{2\to3}^{(1)}(1',3',2')\nonumber\\
&-2\left(\psi_8^{(0)} S_{2''}K_{1r;12}^{(1)} + \psi_8^{(1)} S_{2''}K_{1r;12}^{(0)}\right)   
+K_{3\to3}\psi_8^{(0)} S_{2}   =0 \,.
\label{eq:NLOboot}
\eea
Comparison of Fig.11 with Fig.9 and of (\ref{eq:NLOboot}) with \eqref{eq:LOboot} shows that they are of the same form, except for the new $3\to3$ kernel which appears first in NLO.  
\section{A possible implication for the NLO Odderon solution in $N=4$ SYM}
\setcounter{equation}{0}

The validity of the generalized bootstrap condition (\ref{eq:3bootstrap}) beyond LL might have an interesting 
consequence for the Odderon solution in NLO. 

\subsection{A family of LL Odderon states}

Let us briefly recapitulate the construction of odderon solutions at LL.
The corresponding three reggeized gluon kernel can be written, after colour projection,  as
\bea
H_3&=&\sum_i \omega(q_i)\delta^{(2)}(q'_i-q_i) + K_{8r;12}+ K_{8r;23}+K_{8r;13} \nonumber\\
&=& \frac{1}{2}\left( K_{(12)}+K_{(23)}+K_{(31)} \right),
\label{eq:BKP_LL}
\eea
where $K_{8r;ij} = \frac{N}{2} K_{r;ij}$ is the LL BFKL color octet kernel (including only real emissions) 
acting on gluons with labels $i$ and $j$. 
It is convenient to introduce the infrared safe kernels:
\be
K_{(ij)}= \omega(q_i)\delta^{(2)}(q'_i-q_i)+\omega(q_j)\delta^{(2)}(q'_j-q_j) 
+2 K_{8r;ij}(q'_i,q'_j; q_i,q_j)\,.
\ee
In the following we shall also use $K_{1r;ij}=2 K_{8r;ij}$. Note that in LO this kernel coincides with the color singlet BFKL Hamiltonian $H_2$.  
In each term of the odderon kernel the 'non interacting' gluon trajectory functions are accompanied by delta distributions which will be omitted in the following.
Such a kernel, due to the bootstrap property related to gluon reggeization, satisfies the following relation, when integrated with a constant function (unity),
\be
K_{(ij)}\otimes_{ij} 1=2 \omega(q'_i+q'_j)-\omega(q'_i)-\omega(q'_j)\,,
\ee
where $\otimes_{ij}$ stands for integration in $q_i$ and $q_j$ with the constraint of momentum conservation.

Let us consider the BLV-ansatz \cite{Bartels:1999yt} for the odderon solution (amputated):
\bea
\psi(q_1,q_2,q_3)&=&\varphi(q_1+q_2,q_3)+\varphi(q_2+q_3,q_1)+\varphi(q_3+q_1,q_2)\nonumber\\
&=&\varphi_{12,3}+\varphi_{23,1}+\varphi_{31,2}=
S_{q_1} \otimes_{2} \varphi(q_2,q_3)+{\rm cycl. \ perm.}\,,
\label{eq:ansatz}
\eea
where $\varphi(k_1,k_2)$ denotes a two gluon amplitude. 
Let us act with the odderon kernel on the first term (the other two are obtained by cyclic permutations).
We have three contributions:
\bea
K_{(12)} \otimes_{12} \varphi_{12,3}=\left[ 2 \omega(q'_1+q'_2)-\omega(q'_1)-\omega(q'_2) \right] \varphi(q'_1+q'_2,q'_3),
\label{eq:K-BFKL}
\eea
\be
 K_{(23)} \otimes_{23} \varphi_{12,3}=\left[\omega(q'_2)+\omega(q'_3) \right] \varphi(q'_1\!+\!q'_2,q'_3)+K_{1r;23} \otimes_{23} 
S_{q'_1} \otimes_{2} \varphi(q_2,q_3)
\ee
and
\be
K_{(31)} \otimes_{31} \varphi_{12,3}=\left[\omega(q'_3)+\omega(q'_1) \right] \varphi(q'_1\!+\!q'_2,q'_3)+K_{r;31} \otimes_{23} 
S_{q'_2} \otimes_{1} \varphi(q_1,q_3) \,.
\ee
Summing these last three intermediate expressions one has
\bea
K_{(12)}\otimes \psi^{amp}&=&2 \left[ \omega(q'_1+q'_2) +\omega(q'_3)\right]\varphi(q'_1+q'_2)+\nonumber\\
&{}&K_{1r;23} \otimes_{23} S_{q'_1} \otimes_{2} \varphi(q_2,q_3)+K_{1r;31} \otimes_{23} S_{q'_2} \otimes_{1} \varphi(q_1,q_3)\,.
\eea
We now make use of the relation (\ref{eq:LOboot}) illustrated in Fig. 9 (which was derived as a consequence of the bootstrap equation).
We rewrite it for convenience for this specific case multiplying it by with $N$ (since $N K_{r} = K_{1r}$):
\be
K_{1r;23} \otimes_{23}  S_{q_1} \otimes_{1} + N K_{2 \to 3}(1'2'3') +K_{1r;13} \otimes_{13} S_{q_1} \otimes_{1} + NK_{2\to3} (2'1'3')  =2S_{q'_1} \otimes_{1}  K_{1r;12} \,,
\label{eq:gen-commutator-LO}
\ee
and taking into account the factor $1/2$ in Eq.~\eqref{eq:BKP_LL} we find
\bea
H_3 \otimes \varphi_{12,3}&=&\left[ \omega(q'_1\!+\!q'_2) +\omega(q'_3)\right]\varphi(q'_1\!+\!q'_2)+K_{1r}(q'_1\!+\!q'_2,q'_3;q_1,q_3)\otimes_{13} \varphi(q_1,q_3)\nonumber\\
&{}&-\frac{N}{2}\left[ K_{2\to_3}(1'2'3')+K_{2\to3}(2'1'3') \right] \otimes \varphi\nonumber\\
&=& (H_2 \otimes \varphi)(q'_1\!+\!q'_2,q'_3) -\frac{N}{2}\left[ K_{2\to3}(1'2'3')+K_{2\to3}(2'1'3') \right] \otimes \varphi\,.
\eea

Let us now consider the full ansatz for the solution given in Eq.~\eqref{eq:ansatz}. On applying $H_3$ we get
\bea
H_3 \psi&=&H_3 \otimes (\varphi_{12,3}+\varphi_{23,1}+\varphi_{31,2})\nonumber\\
&=& (H_2 \otimes \varphi)_{12,3}+(H_2 \otimes \varphi)_{23,1}+(H_2 \otimes \varphi)_{31,2}\nonumber\\
&&-\frac{N}{2}\left[ K_{2\to3}(1'2'3')+K_{2\to3}(2'1'3') +K_{2\to3}(2'3'1')
+K_{2\to3}(3'2'1')\right. \nonumber\\ &&\left. +K_{2\to3}(3'1'2') + K_{2\to3}(1'3'2') \right] \otimes \varphi.
\eea

In \cite{Bartels:1999yt} it was observed that if one choses, in the ansatz (\ref{eq:ansatz}), a function $\varphi(k_1,k_2)$ which is antisymmetric under the exchange of $k_1$ and $k_2$ then, thanks to the symmetry properties of $K_{2 \to 3}$, the sum of terms in the last two lines is zero and
\be
H_3 \psi = \sum_{cycl.perm.} \left(\psi_8 H_2 \varphi \right)_{12,3}.
\ee
Moreover if $\varphi$ is taken as an antisymmetric eigenfunction $\varphi_h$ of the singlet BFKL kernel $K_{1r}$ 
with eigenvalue $\chi_h$,
then $\psi^{amp}$ becomes an eigenfunction of the odderon hamiltonian:
\be
H_3 \, \psi_h=\chi_h \, \psi^{amp}_h.
\ee
It is well-known that the antisymmetric BFKL eigenfunctions have odd conformal spin quantum numbers  $n$, and the 
leading eigenvalue $\chi(\nu,n)$ belongs to $\nu=0$ and $n=1$ with:
\be
\chi(0,0) = 0,
\ee 
i.e. the intercept of this Odderon solution equals zero. 

It is important to emphasize that a crucial 
ingredient in this argument was the bootstrap identity (\ref{eq:gen-commutator-LO}) and the symmetry of 
the kernel $K_{2\to3}(q_1,q_2;q'_1,q'_2,q'_3)$ under the exchange of momenta: $q_1\leftrightarrow q_2$
and $q'_1 \leftrightarrow q'_3$.   

\subsection{An ansatz for a family of $N=4$ SYM Odderon states at NLO}

Let us now go to NLO and assume that the bootstrap condition (\ref{eq:3bootstrap}) 
holds in NLO, in particular for the signature configuration $(\tau_1,\tau_2)=(+,-)$ with the $t_1$-channel being 
in the color singlet configuration.  As we have argued above, the validity of (\ref{eq:3bootstrap}) implies 
Fig.11, i.e. a generalized version of the commutator relation (\ref{eq:NLOboot}) also holds in NLO. 
From now on we specialize to the case of $N=4$ SYM where special simplifications occur.

In \cite{Bartels:2012sw} the NLO corrections for the Odderon kernel have been calculated. They consist of the 
new $3 \to3$ kernel and the NLO corrections for the symmetric BFKL kernel in the color octet state. 
In particular, it was found that, in the supersymmetric case, due to the fact that also the scalar and fermion fields belong to the adjoint representation of the gauge group,
the symmetric ($d_{abc}$) and antisymmetric ($f_{abc}$) NLO corrections for the BFKL kernel  coincide.
This implies that the NLO wave function for the reggeized gluon coincides with the NLO wave function for the symmetric $d$-reggeon: the degeneracy between the odd signature $f$-reggeon (gluon) and the even signature $d$-reggeon remains valid in NLO.  We note that this degeneracy is not true for QCD in general, but it does apply for the particular case of QCD in the planar (large $N_c$ limit) . Moreover the $3\to3$ connected kernel recently computed in the $d$-projected color singlet channel is the same for both N=4 SYM and QCD. This implies that the NLO generalization of the commutator relation (\ref{eq:commutator}) remains the same if we replace the overall $f$ structure by an overall $d$ structure, and the $f$-reggeon by the $d$ reggeon. 

It is worthwhile to mention a specific feature of the NLO results obtained in  \cite{Bartels:2012sw}. For the calculation of the $3 \to 3$ kernel it was important to observe that the s-channel intermediate state gluons, initially, had to be taken as being off-shell. This lead to ultraviolet divergencies of the integrals over longitudinal variables which cancelled only  
after summing over all permutations of reggeized gluons. In the final result, i.e., after taking the sum over all permutations, the  $3 \to 3$ kernel could be written as a sum of three cyclic permutations, each of them being 
the residue of the two poles of the s-channel gluon propagators,
with the special constraint that the longitudinal momenta of the s-channel gluons had opposite directions. It is this feature 
which allows to identify the $3 \to 3$ vertex inside the double energy discontinuity (Fig.10) with the $3 \to 3$ vertex 
derived in  \cite{Bartels:2012sw}.    

With these arguments, we now show that all the ingredients which went into the LO solution of the 
Odderon eigenvalue problem can be extended to NLO. As a result, the Odderon solution, again, is given by the $n=1$ color singlet BFKL eigenfunction with intercept zero.  

Let us start from the structure of the NLL BKP kernel~\cite{Bartels:2012sw} in the Odderon channel, i.e. having taken the projection on the $d_{a_1a_2a_3}$ color singlet structure for the $N=4$ SYM theory. The Hamiltonian now has 
the form:
\bea
H_3&=&\omega_1+\omega_2+\omega_3+K_{8r;12}+K_{8r;23}+K_{8r;31}+K_{(123)}\nonumber\\
&=&\frac{1}{2} \left( K_{(12)}+K_{(23)}+K_{(31)} \right)+K_{(123)}\,,
\eea
where $\omega_i$ are the reggeized gluon trajectories, $K_{8r;ij}$ is the real part of the BFKL kernel in the symmetric octect channel, all of them now up to NLO accuracy, and $K_{(123)}$ is the connected three-gluon interaction term (in the tree approximation) which contributes first in NLO. The kernels in the $8_S$ representation, coinciding in $N=4$ SYM with the ones in the $8_A$ representation are
\bea
K_{(ij)}=\left(\omega_i+\omega_j\right)+2K_{8r;ij}=2 H^{(8)}_{(ij)}-\left(\omega_i+\omega_j\right)
\label{NLOK-ij}
\eea
and together with $K_{(123)}$ (also the same for the $f_{abc}$ and $d_{abc}$ singlet color states) are associated to infrared finite operators. 
It will be convenient to make also use of the color singlet Hamiltonian $H_2$:
\be
H_2 = \omega(q_1) + \omega(q_2) + K_{1r;ij}.
\ee
Since in NLO the real part of the BFKL singlet kernel, $K_{1r}^{(1)}$, is different from $2 K_{8r}^{(1)}$, the NLO part 
of $H_2$ is different from that of $K_{(ij)}$.

An important property of the octet kernel, valid also in $N=4$ SYM, is the strong bootstrap equation
\bea
H^{(8)}_{(ij)}\psi_8(q_i,q_j)=\omega(q_i+q_j)\psi_8(q_i,q_j) \,,
\label{NLO_strong_boot}
\eea
where $\psi_8$ is the gluon wave function
\bea
\psi_8=\psi_8^{(0)}+\psi_8^{(1)} \eea
and which can be translated in the equivalent relation for the IR safe kernel
\be
K_{(ij)} \psi_8(q_i,q_j)=\left[2\omega(q_i+q_j)-\omega(q_i)-\omega(q_j)\right] \psi_8(q_i,q_j) \,.
\label{eq:NLOstrongbootstrap1}
\ee
Upon perturbative expansion to NLO accuracy, it becomes:
\bea
\!\!\!\!\!\!K^{(0)}_{(ij)} \psi_8^{(1)}(q_i,q_j)+K^{(1)}_{(ij)} \psi_8^{(0)}(q_i,q_j)&\!\!\!=&\!\!\!\!
\left[2\omega^{(0)}(q_i+q_j)\!-\!\omega^{(0)}(q_i)\!-\!\omega^{(0)}(q_j)\right] \psi_8^{(1)}(q_i,q_j) \nonumber\\
&+&\!\!\!\!\left[2\omega^{(1)}(q_i+q_j)\!-\!\omega^{(1)}(q_i)\!-\!\omega^{(1)}(q_j)\right] \psi_8^{(0)}(q_i,q_j)\,.
\label{eq:NLOstrongbootstrap2}
\eea
Let us now consider an ansatz which extends the LL BLV wavefunction  $\psi^{(0)}$ solution reviewed above to NLO:
\be
\psi=\psi^{(0)}+\psi^{(1)} \,.
\ee
We again consider a two gluon wave function  $\varphi= \varphi^{(0)}+ \varphi^{(1)}$ (which later on will be chosen to be antisymmetric)
and define
\bea
\psi(q_1,q_2,q_3)= \!\!\!\sum_{cycl. \ perm.} \!\!\! \psi_8(q_1,q_2;q_1+q_2)\varphi(q_1+q_2,q_3)
=\!\!\!\sum_{cycl. \ perm.} \!\!\! \psi_8\, \hat{S}_{q_1}\varphi\,.
\label{eq:genans}
\eea
Note that the notation $\psi_8\, \hat{S}_{q_1}$ stands for a splitting of a gluon carrying momentum $q_1$,  
multiplied by  the eigenfunction $\psi^{(8)}$ solution of (\ref{NLO_strong_boot}). In more detail we have, for the first of the three cyclic permutations  
up to NLL accuracy:
\bea
\psi^{(1)} = \!\!\!\sum_{cycl. \ perm.} \!\!\left[ \psi_8^{(0)} \varphi^{(1)}(q_1+q_2,q_3)+
 \psi_8^{(1)}(q_1,q_2;q_1+q_2) \varphi^{(0)}(q_1+q_2,q_3)\right]\,.
\eea

Under which conditions can the wave function defined in Eq.~\eqref{eq:genans} can be an eigenstate of the odderon kernel $H_3$? To answer this we proceed in the same way as in the provious subsection, i.e. we compute $H_3 \psi$ and 
show that, with an antisymmetric function $\varphi$, $\psi$ is, in fact, eigenfunction of $H_3$ with the kernel $H_2$.   
%

Expanding up to NLO let us consider the following notation: $\omega_i=\omega^{(0)}_i+\omega^{(1)}_i$, 
$K_{8r(ij)}=K^{(0)}_{8r(ij)}+K^{(1)}_{8r(ij)}$, $K_{(ij)}=K^{(0)}_{(ij)}+K^{(1)}_{(ij)}$, $H_2=H_2^{(0)}+H_2^{(1)}$,
and $H_3=H_3^{(0)}+H_3^{(1)}$. We wish to compute 
\bea
H_3 \psi =
\left( \frac{1}{2}\left[ K_{(12)}+K_{(23)}+K_{(31)}\right]+K_{(123)}\right)\!\!\! \sum_{cycl. \ perm.} \!\!\! \psi_8\, \hat{S}_{q_1} \varphi
\eea
which, up to NLO, means:
\be
\left( H_3 \psi \right)^{(1)}= \sum_{cycl.perm.} \left(H_3^{(1)} \psi_8^{(0)} \varphi_{12,3}^{(0)}+  H_3^{(0)} \psi_8^{(1)} \varphi_{12,3}^{(0)}
+ H_3^{(0)} \psi_8^{(0)} \varphi_{12,3}^{(1)} \right)\,.
\label{eq:H3action}
\ee
The last term containing $\varphi^{(1)}$ can be treated in the same way as the LO solution: if $\varphi^{(1)}$ is 
an antisymmetric function, this last term is eigenfunction by itself:
\be
H_3^{(0)} \sum_{cycl.perm.}  \psi_8^{(0)} \varphi_{12,3}^{(1)} = \sum_{cycl.perm.} (\psi_8^{(0)}H_2^{(0)} \varphi^{(1)})_{12,3} \,.
\label{eq:phi0}
\ee

In the first two terms of (\ref{eq:H3action}) we begin with the first of the three cyclic permutations, $\varphi_{12,3}$, and apply the kernel 
$K_{(12)}$: this is just (\ref{eq:NLOstrongbootstrap1}), i.e. we have the NLO term of  the strong bootstrap condition (\ref{eq:NLOstrongbootstrap2}). 
They can be written as
\be
\frac{1}{2}\left( 2\omega_{12}^{(1)}-\omega_{1}^{(1)}-\omega_{2}^{(1)}\right) \psi_8^{(0)} S_{q_1} \otimes \varphi^{(0)}+
\frac{1}{2}\left( 2\omega_{12}^{(0)}-\omega_{1}^{(0)}-\omega_{2}^{(0)}\right) \psi_8^{(1)} S_{q_1} \otimes \varphi^{(0)}\,.
\label{eq:bootterm}
\ee
Next we consider the action of the remaining parts of $H_3$ on  $\varphi_{12,3}$: 
\bea
\frac{1}{2}\left[ K^{(1)}_{(23)}+K^{(1)}_{(31)}\right] \otimes \psi_8^{(0)}S_{q_1} \otimes \varphi^{(0)}
+\frac{1}{2}\left[ K^{(0)}_{(23)}+K^{(0)}_{(31)}\right] \otimes \psi_8^{(1)} S_{q_1}\otimes \varphi^{(0)}+\nonumber\\
K_{(123)} \otimes \psi_8^{(0)} S_{q_1}\otimes \varphi^{(0)}. 
\eea
Using (\ref{NLOK-ij}) for the kernels $K_{ij}$ this expression can be rewritten as
\bea
\frac{1}{2} \left( \omega_1^{(0)}+\omega_2^{(0)}+2 \omega_3^{(0)}\right)\psi_8^{(1)}S_{q_1} \otimes \varphi^{(0)} +\frac{1}{2}\left( \omega_1^{(1)}+\omega_2^{(1)}+2 \omega_3^{(1)}\right)\psi_8^{(0)}S_{q_1} \otimes \varphi^{(0)} \nonumber\\
+\left( K^{(1)}_{8r;23}+K^{(1)}_{8r;31}\right) \otimes \psi_8^{(0)}S_{q_1} \otimes \varphi^{(0)}
+\left( K^{(0)}_{8r;23}+K^{(0)}_{8r;31}\right) \otimes \psi_8^{(1)} S_{q_1}\otimes \varphi^{(0)}+\nonumber\\
+K_{(123)} \otimes \psi_8^{(0)} S_{q_1}\otimes \varphi^{(0)} \Bigr) \hspace{1cm}\,.
\label{eq:equality0}
\eea 
We now use, for the last two lines of this equation,  the relation (\ref{eq:NLOboot}) (or Fig.11) (which was derived from the validity of the generalized bootstrap condition) and obtain for this part of (\ref{eq:equality0}):
\be
2 \psi_8^{(0)} S_{q_1'}\otimes K_{1r}^{(1)}\otimes \varphi^{(0)}_{12,3}+
2\psi_8^{(1)} S_{q_1'}\otimes K_{1r}^{(0)}\otimes \varphi^{(0)}_{12,3}
- K^{(1)}_{2\to3}(1'2'3') \otimes \varphi^{(0)}-  K^{(1)}_{2\to3}(2'1'3')\otimes \varphi^{(0)}\,.
\label{eq:transf}
\ee
Let us now follow the argument made for the derivation of the LO solution and take $\varphi^{(0)}$ to be an antisymmetric function. Then, after summing 
over all cyclic permutations and making use of the symmetric properties of the $2\to 3$ kernel, the sum of terms 
containing  $K^{(1)}_{2\to3}$ cancels.
Collecting all our results in Eqs. (\ref{eq:bootterm}), (\ref{eq:equality0}) and (\ref{eq:transf}) we find for the first two terms of (\ref{eq:H3action}):      
\bea
\sum_{cycl. \ perm.} \!\!\! \Bigl(
\left[ \omega^{(0)}_{12} + \omega^{(0)}_3 \right] \psi_8^{(1)} \ S_{q_1}\otimes \varphi^{(0)}+
\left[\omega_{12}^{(1)} + \omega^{(1)}_3 \right]\psi_8^{(0)} \  S_{q_1}\otimes \varphi^{(0)} + \nonumber\\
\psi_8^{(0)}  S_{q_1'}\otimes K_{1r}^{(1)}\otimes \varphi^{(0)}+ \psi_8^{(1)} S_{q_1'}\otimes K_{1r}^{(0)}\otimes \varphi^{(0)}
\Bigr)
\nonumber\\
= \sum_{cycl. \ perm.}  \left( (\psi_8^{(1)} H_2^{(0)}+ \psi_8^{(0)} H_2^{(1)})  \varphi^{(0)} \right)_{12,3}\,.
\eea
Combining  this with (\ref{eq:phi0}), we have thus verified that, in NLO,
\be
H_3 \psi = \sum_{cycl. \ perm.} \left(\psi_8 H_2 \varphi \right)_{12,3}.
\ee
Finally, if up to NLO, $\varphi$ is an odd eigenfunction of $H_2$ with the NLO eigenvalue $\chi(\nu,n)=\chi^{(0)}(\nu,n) + \chi^{(1)}(\nu,n)$,
our function $\psi$ is eigenfunction of the odderon Hamiltonian with the leading eigenvalue  
\be
\chi(\nu=0,n=1) = 0.
\ee

\section{Conclusions}
In this paper we have shown that there exist, at least in the leading logarithmic approximation, generalized 
bootstrap relations (\ref{eq:3bootstrap}) which are essential to guarantee the self-consistency of the ansatz (\ref{eq:decomp}),
in particular the compatibility between single and double energy discontinuities. They can be seen as a generalization 
of the familiar bootstrap equation (\ref{eq:2bootstrap}) which guarantees, for the $2 \to 2$ scattering amplitude with 
color octet exchange, the consistency between the leading real part and the energy discontinuity (imaginary part).
For this case the bootstrap condition has been shown to remain valid also in NLO accuracy. It is therefore 
tempting to expect that also the generalized version of the bootstrap equation discussed in this note is valid 
in NLO.       

If these generalized bootstrap relations are valid in NLO, they allow an interesting application for the 
NLO Odderon solution in $N=4$ SYM. After reviewing the derivation of the LO Odderon solution we have shown that the same line of arguments can be applied also in NLO, i.e. there exists a family of NLO eigenfunctions which are constructed with the $n=1$ color singlet BFKL eigenfunctions and share the same intercept up to one. In our argument, however, we are making use of a few properties of NLO corrections which are valid in $N=4$ SYM but not in QCD. Therefore, without further direct QCD calculations we cannot conclude that also in QCD, for finite $N_c$, the NLO Odderon  has intercept one.  Nevertheless it is possibile to make a weaker statement, that is that at NLO for QCD in the planar limit (large $N_c$) the Odderon exchange has a leading intercept at one.  

It is clear that a direct proof of the NLO bootstrap equation (\ref{eq:3bootstrap}) is of vital interest.
\vskip 0.5cm

\noindent
{\bf Aknowledgements}\\
One of us, J. B., gratefully acknowledges the hospitality of the INFN Section of Bologna where most of this work has been done.


\end{document}